# *Looking for new thermoelectric materials among TMX intermetallics using high-throughput calculations*

Celine Barreteau, Jean-Claude Crivello, Jean-Marc Joubert, Eric Alleno

Université Paris Est, ICMPE (UMR 7182), CNRS, UPEC, F-94320 Thiais, France

## Highlights

- Search for new thermoelectric intermetallic compounds
- Screening via high-throughput calculations of 2280 configurations of *TMX* compounds
- Identification of 21 possible stable and semiconducting *TMX* compounds

## Abstract

Within 4 different crystal structures, 2280 ternary intermetallic configurations have been investigated via high-throughput density functional theory calculations in order to discover new semiconducting materials. The screening is restricted to intermetallics with the equimolar composition *TMX*, where *T* is a transition metal from the Ti, V, Cr columns, Sr, Ba, Y and La, *M* an element from the first line of transition metals and *X* a *sp* elements (Al, P, Si, Sn and Sb), i.e. to a list of 24 possible elements. Since the calculations are done combinatorically, every possible ternary composition is considered, even those not reported in the literature. All these *TMX* configurations are investigated in the 4 most reported structure-types: TiNiSi, MgAgAs, BeZrSi and ZrNiAl. With an excellent agreement between calculations and literature for the reported stable phases, we identify 472 possible stable compounds among which 21 are predicted as non-metallic. Among these 21 compositions, 4 could be considered as new semiconductors.

## I. Introduction

Thermoelectric devices can be used as refrigerators or as electric power generators making them interesting materials which may play a role in sustainable development. Their conversion efficiency is related to the dimensionless figure of merit *ZT* of their constituting materials [1]. *ZT* is defined by the expression $ZT = (\alpha^2/\rho)(T/\lambda)$ with α the Seebeck coefficient, ρ the electrical resistivity, λ the thermal conductivity and *T* the average temperature. Well-established materials are already used in commercial devices such as $Bi_2Te_3$ [2] in thermoelectric refrigerators or Si-Ge [3] and PbTe [4] in thermoelectric generators. However their low performance as well as their cost (Ge 1000$/kg or Te 150$/kg [5]) restrict them to niche markets. Even though other interesting families have recently emerged, there are still limitations regarding their figure of merit or their stability. Thus, research for new thermoelectric material is required and, more specifically, new thermoelectric materials which display a moderate cost.

In recent years, theoretical approaches integrating first-principles calculations or machine learning have constantly progressed. Indeed, these methods allow to screen large set of compounds relying on always more efficient computer codes and faster computers. By using compounds databases such as the Inorganic Crystal Structure Database (ICSD) [6] or by selecting a very large number of them to start with, AFLOWLIB [7, 8], Materials Project [9] or "Thermoelectrics Design Lab" [10] respectively, screened them for specific applications. Indeed, it is possible to study in a much shorter time than experimentally thousands of compounds or structures. These methods hence allow the identification of new compounds [11], the prediction of their thermodynamic stability [12, 13] or the discovery of compounds displaying promising specific properties [14-16].

In this article, we investigate thousands of ternary intermetallic compounds via high-throughput calculations in order to discover new semiconducting materials. We restrict our screening to intermetallics with the equimolar composition *TMX,* with *T* a transition metal from the Ti, V, Cr columns, Sr, Ba, Y and La, *M* an element from the first line of transition metals and *X* a *sp* elements (Al, P, Si, Sn and Sb). The calculations are done combinatorically, therefore every possible ternary composition is calculated in the four most reported structure-type: TiNiSi, MgAgAs, BeZrSi and ZrNiAl. The MgAgAs structure-type corresponds to the family of the half-Heusler alloys which contains several already known thermoelectric materials [17-19]. This choice of crystal structures thus enlarges the search of new semi-conducting compounds within those displaying the 1:1:1 stoichiometry. Similarly, Gautier *et al.* [13] determined the thermodynamical stability and electronic structure of only unreported 18-electron *TMX* compounds. We did not restrict our calculations to 18-electron compositions and considered as well the reported compounds. The list of chemical elements is restricted to those that are not too rare or too expensive, in order not to preclude some applications.

The calculations are based on the Density Functional Theory (DFT), which allows us to calculate the enthalpy of formation ($\Delta_f H$) of each compound as well as their density of states as a function of energy. In a first step, the most stable of the four calculated structures for each composition is compared with literature, which allows validating the robustness of our calculations. In a second step, the density of states of the selected compounds is analysed in order to identify potential non-metallic (semi-conducting or semi-metallic) compounds for thermoelectric application.

## II. Calculations and Methodology

A combinatorial approach is implemented in order to screen all possible ternary combinations generated within a restrained set of chemical elements. Indeed, in order to find cheap thermoelectric materials, the screening is carried out on ternary intermetallics *T-M-X*, with *T*, *M* and *X* which do not span the entire periodic table [10, 20] but a well-defined set of elements. The toxic, rare or expensive elements have been excluded, even though they are found in several thermoelectric compounds as this can later be an obstacle to the development of bulk thermoelectric devices. *T* is a transition metal from the Ti, V, and Cr columns or Sr, Ba, Y and La, *M* an element from the first line of transition metals and *X* a *sp* element (Al, P, Si, Sn and Sb). This set includes heavy elements, favourable to a low thermal conductivity, such as Ta, W, Ba, La, Sn or Sb cheaper than Ge or Te. The full list of included elements is displayed in the periodic table of the elements in Figure 1.

Once the 15 binary combinations with *T* = *M* excluded, 570 (13 *T* x 9 *M* x 5*X* - 15) stoichiometric 1:1:1 ternary compositions are to be investigated. A preliminary examination of the crystal databases [21] allowed us to reference 188 known and stable compounds among all the 570 possible *TMX* compositions. For the other 382 possible combinations, there is no report of their synthesis and characterization in the databases, meaning that they are either unknown or unstable for the 1:1:1 stoichiometry. All the 188 reported compounds crystallize in 13 different crystal structure-types. In Pearson's [21], not only 13 but 41 structure-types are reported for the 1:1:1 stoichiometry owing to a set of elements spanning the entire periodic table [13, 22]. To further limit the amount of calculations, only the structure-types with more than 10 reported compounds were kept: the orthorhombic TiNiSi (*Pnma*), the cubic MgAgAs (*F*-43*m*, half-Heusler) and the two hexagonal ZrNiAl (*P*-62*m*) and BeZrSi ($P6_3/mmc$), presented in the Figure 1. This choice will be a posteriori supported by the results. The 570 possible compositions are calculated in these 4 structure-types by systematically ascribing to each crystallographic site a unique element of our *T*-*M*-*X* nomenclature, discarding other configurations or site mixing (see Figure 1), thus yielding 2280 different configurations.

*Table 1 : Assignment of the TMX set of chemical elements to their crystallographic positions.*

| Structure-type of *TMX* / Wyckoff positions | TiNiSi *Pnma* (62) | ZrNiAl *P*-62*m* (189) | MgAgAs *F*-43*m* (216) | BeZrSi $P6_3/mmc$ (194) |
|---|---|---|---|---|
| *T* | **4*c*** | **3*f*** | **4*b*** | **2*a*** |
| *M* | **4*c*** | **2*d*, 1*a*** | **4*a*** | **2*d*** |
| *X* | **4*c*** | **3*g*** | **4*c*** | **2*c*** |

This method allows to explore a reasonably sized set of new configurations, nonetheless 10 times larger than the 188 known compounds, which can contain new stable compounds. This extended set is also justified by the existing compounds, which are referenced in literature in more than one structure-type.

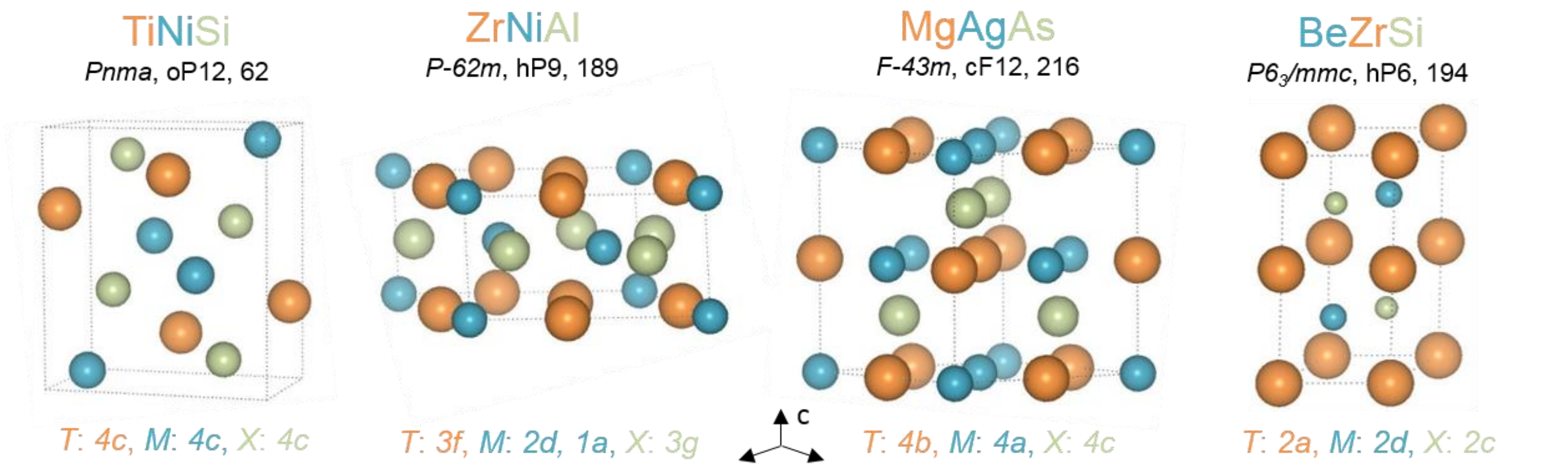

*Figure 1: Unit cell of the four calculated crystal structure-type: TiNiSi, ZrNiAl, MgAgAs and BeZrSi. The T, M and X elements are explicitly assigned to a Wyckoff position, using the following colour code: T atoms are represented by orange balls, M atoms by blue balls and X atoms by green balls.*

The calculations of the stability and ground state properties are based on the DFT, which allows to obtain for each compound the enthalpy of formation at 0 K (obtained by total energies difference compared to the elemental reference state) as well as its electronic structure. They were conducted using the projector augmented wave (PAW) method implemented in the Vienna *ab initio* Simulation Package (VASP) [23-27]. The exchange correlation was described by the generalized gradient approximation modified by Perdew, Burke and Ernzerhof (GGA-PBE) [28]. Energy bands up to a cut off energy E = 600 eV were used in all calculations. A high density *k*-points meshing was employed for Brillouin zone integrations in the TiNiSi structure-type (7 × 11 × 6), MgAgAs (21 × 21 × 21), ZrNiAl (19 × 19 × 37) and BeZrSi (21 × 21 × 10). These parameters ensure good convergences for the total energy. The convergence tolerance for the calculations was selected as a difference on the total energy within 1 x $10^{-6}$ eV/atom. We performed for each structure type, volume and ionic relaxation steps and we considered magnetism for all the configurations. Blöchl correction was considered at a final step calculation [29].

To screen the calculated output data, several criteria were implemented in our approach. The first criterion is related to the thermodynamic stability of the compound. We excluded those with a positive enthalpy of formation. Indeed, a positive enthalpy of formation means that a configuration is not stable against the decomposition into the pure elements. However, even if a negative enthalpy of formation is necessary, it is not a sufficient condition for the stability of a compound at 0 K. The most negative enthalpy of formation of the four competing structure-types at the same composition allows to select the most probable stable structure. Finally, the comparison of these results with the reported data allowed to validate the robustness of the model by checking if the calculated most stable structure-type corresponds to the reported one in the literature.

In a second step, the electronic structures of the stable configurations were analysed. Indeed, compounds that exhibit a density of states at the Fermi level higher than 0.5 states/eV by atom were considered as metallic and excluded. The criterion was not defined at zero in order to integrate the incertitude on the calculation and more specifically the one on the determination of the energy gap within the GGA-PBE approximation.

Both criteria, stability and electronic structure allowed to screen the set of configurations and led to a restricted set which corresponds to potentially stable semiconductors compounds. The last step of this work will focus on the specific study of these 21 potential semiconductors compounds.

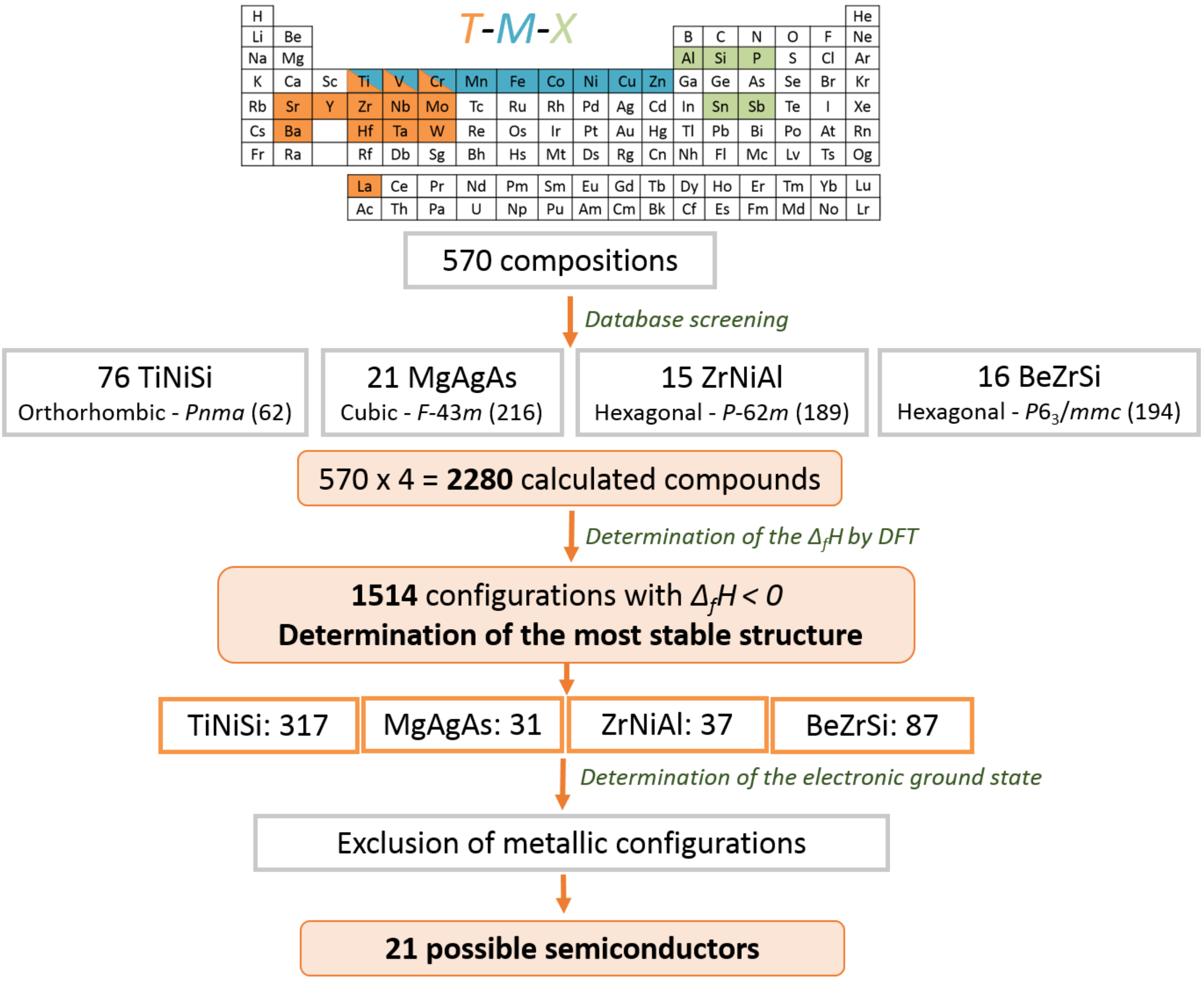

*Figure 2 : Overview of the screening method: complete workflow of the method. The periodic table in the upper part presents elements included in the chosen set: T in orange corresponds to the transition metals from the Ti, V and Cr column, Sr, Ba, Y and La; M in blue corresponds to the first line of transition metals and X in green corresponds to the sp elements. For each structure-type, the number of referenced compounds are indicated.*

## III. Results

Results of the screening are presented in the Figure 3. For each *X*, the possible *T-M* combinations are listed in the X-axis and associated to its most stable structure-type. In the figure, two main information are associated to each configuration: the agreement between the reported data and the calculations in predicting the stable structure-type as well as the nature of the electronic ground state.

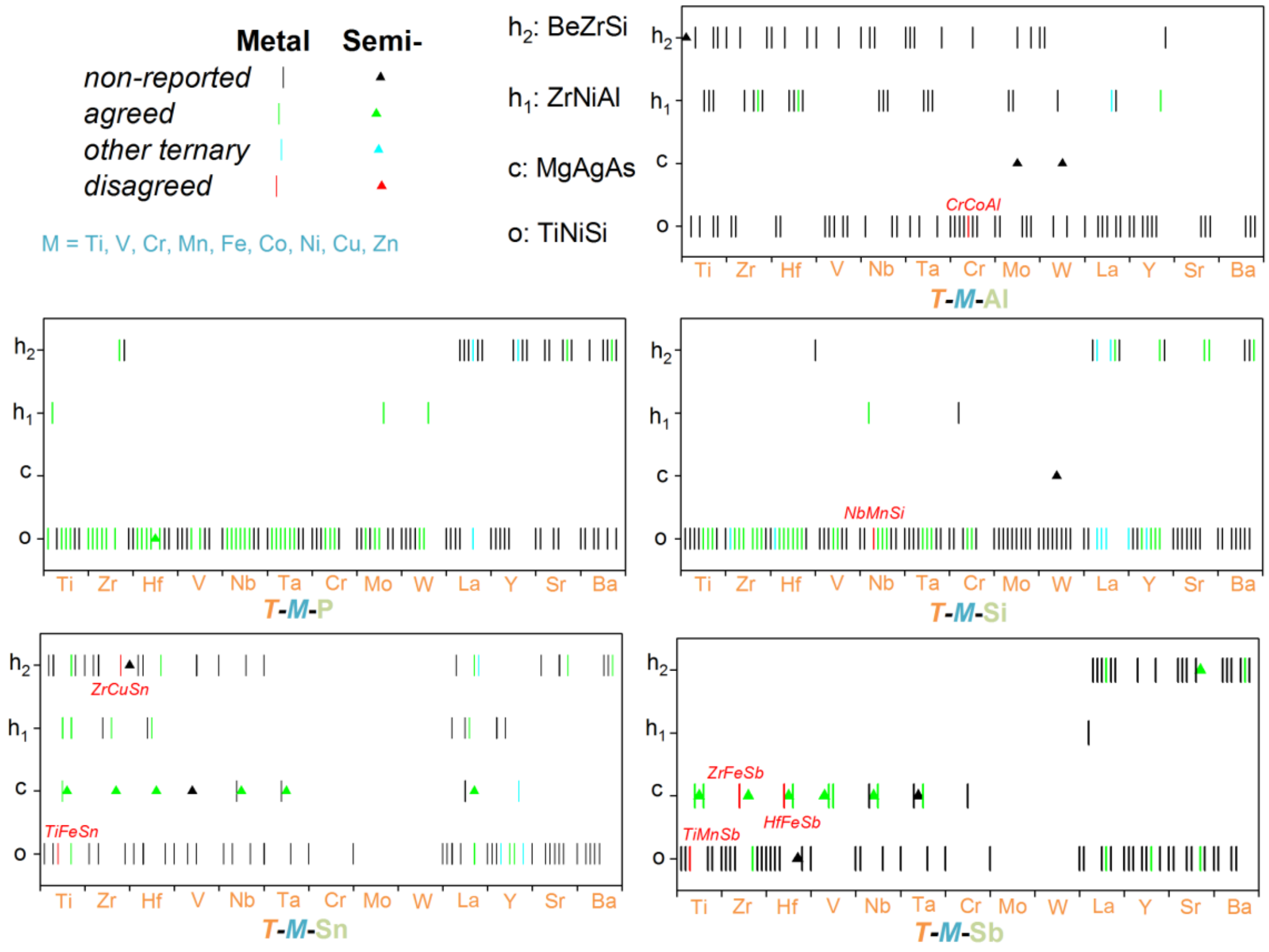

*Figure 3: Plot of the calculated T-M-X configurations in their most stable structure-type; matching between calculated stable structure-type and reported data are defined by colour: black is for compounds which are not reported in the crystal databases, green represents the agreement, red is disagreement and blue is for compound reported in structure-type not included in our calculations. Vertical line corresponds to metallic compounds and triangle to semiconductors. For each T element, the 9 M elements are listed in the same order as in the periodic table (Ti, V, Cr …).*

The calculations allow to exclude 98 from the 570 possible compositions which do not present a negative enthalpy in any of the 4 prototypes, these compounds are not reported in figure 3. These unstable compounds correspond essentially to the *T-M*-Sn or *T-M*-Sb when *T* is from the Cr column. Indeed, there is no W-M-Sn/Sb and only two Mo-M-Sn/Sb calculated as stable in the 4 structure-types. In the remaining 472 possible compounds, the TiNiSi structure-type is the richest structure-type with 2/3 of the compositions contained (317 compounds stable in this prototype). The hexagonal types are less represented with 87 compositions in the BeZrSi structure-type and 37 in the ZrNiAl structure-type. Finally, the half-Heusler or MgAgAs types count 31 stable compositions. Taking into account uncertainties in the calculations, several structures can be attributed to more than one structure-type as the difference in their $\Delta_f H$ value is lower than 0.5 kJ/mol. For these 34 compositions, the 2 or 3 possible structure-types are represented in the Figure 3. As mentioned before in this paper, to reduce the number of calculations, only the most common structure-types were chosen. However for 16 of the 472 possible compositions, other ternary structure-types were reported in the crystal databases. To distinguish them from the other results, they are represented by a specific colour in Figure 3.

Trends can also be noticed between structure-type and nature of the *X* element. Indeed, the MgAgAs structure-type is only found for compositions where *X* = Sn and Sb, except for 3 unreported compounds: MoCoAl, WCoAl and WFeSi. On the other hand, for *X* = P or Si, most of the compositions crystallize in the orthorhombic structure-type, especially when *T* is a transition metal. Finally, the *TM*-Al and *TM*-Sn compositions mainly crystallize in the hexagonal structure-types.

The direct comparison between crystallographic databases and the results shows a good agreement of 94%, with only 7 compounds which present a disagreement between the predicted and observed crystal structures. These differences between the reported and the calculated structure-types will be discussed later in this paper.

Figure 4 gives another important information for each calculated compound, related to the value of the density of states at the Fermi level. Depending on these values, the compounds are classified as metal or semiconductor. The results clearly show that the metallic ground state is predominant. Only 21 compounds are predicted as non-metallic, among which 50% predicted stable in the cubic MgAgAs structure-type. Some of them correspond to unreported compounds. They will be carefully described in the last part of this article.

As examples, the density of states of 4 existing compounds are displayed in Figure 4. They have been chosen to illustrate our criterion in the definition of metallic or non-metallic state, according to potential error from the exchange correlation functional. Indeed, ZrCoSb and NbCoSi are 2 clear cases: the Fermi level falls in a valley of no density of states for ZrCoSb leading to a semiconducting ground state [30-32], whereas NbCoSi presents a metallic ground state.

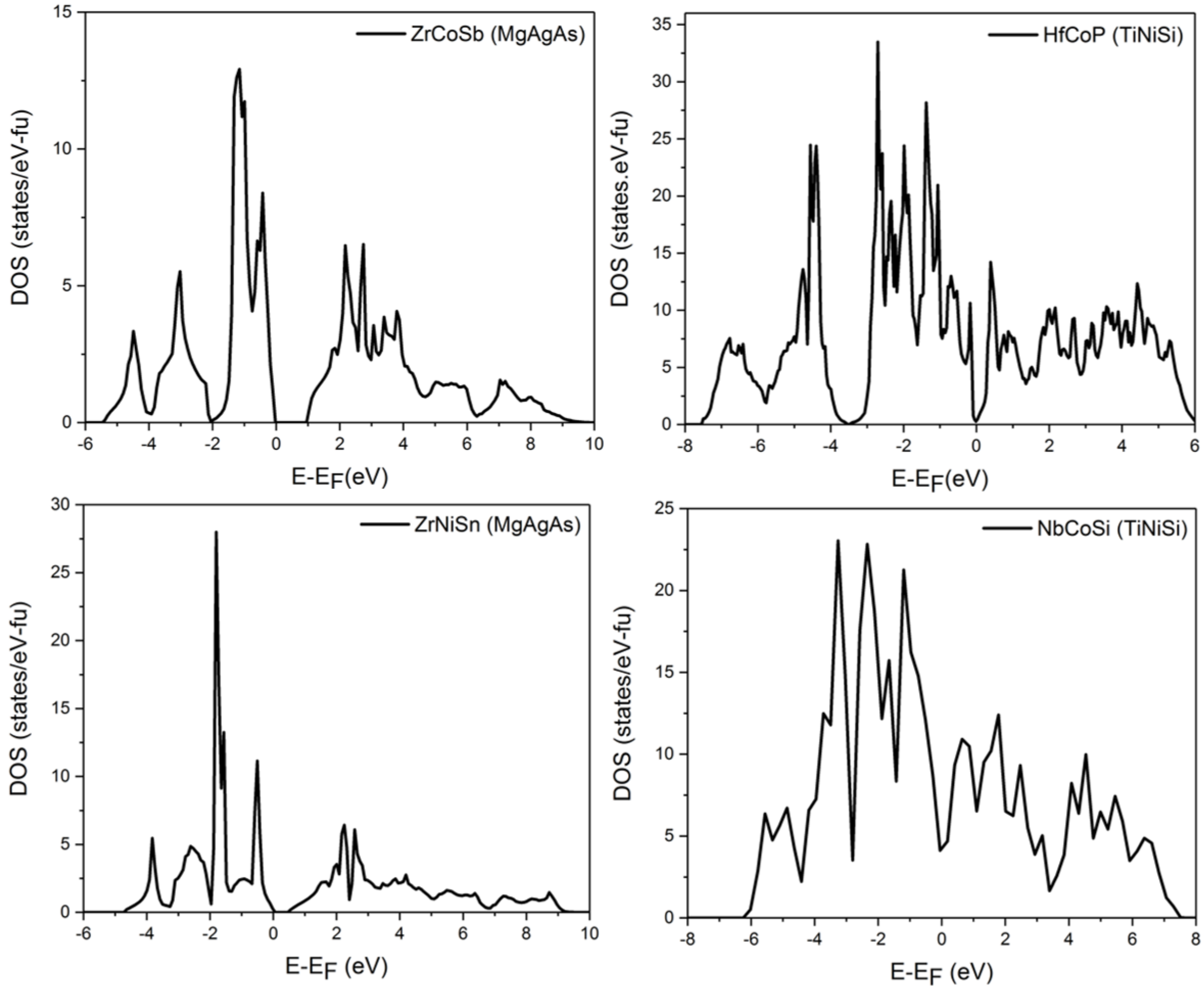

*Figure 4 : Density of states as a function of energy for HfCoP and NbCoSi in the TiNiSi structure-type and ZrCoSb and ZrNiSn in the MgAgAs structure-type.*

ZrNiSn and HfCoP are also classified as potential semiconductors even if their density of states (DOS) is not strictly equal to 0 states/eV-f.u. at the Fermi level (DOS(ZrNiSn) = 0.22 states/eV-f.u. and D.O.S(HfCoP) = 0.38 states/eV-f.u.). In particular, ZrNiSn illustrates the need of flexible criterion as its DOS differs from 0 at the Fermi level while it is reported [33-35] as a semiconductor.

## IV. Discussion

As mentioned before, our method allows the screening of thousands of possible configurations in an efficient way and the agreement between the predicted and reported crystal structures is excellent. Nonetheless, there are 7 compounds (ZrFeSb, HfFeSb, TiMnSb, CrCoAl, TiFeSn, ZrCuSn, and NbMnSi) for which the calculated stable structure-type differs from that reported in the crystallographic databases. It is thus essential to understand the origin of these discrepancies in order to better trust the predictions obtained for the compounds not previously reported in the databases.

### i. Thermodynamic stability

ZrFeSb and HfFeSb are predicted to be more stable in the MgAgAs structure-type whereas they are reported [36] as crystallizing in the TiNiSi structure-type. For both compounds, our calculated $\Delta_f H$ at 0 K differs by a few kJ/mol between these two structure-types. However, even if these compounds are reported as crystallizing in the TiNiSi structure-type, a significant deviation from the full stoichiometry with defect in the *M*-sites, has been noticed in the literature [36, 37]. Indeed, there is no report of the synthesis of stoichiometric ZrFeSb or HfFeSb. This deviation could explain the difference between the calculations and the experimental reports, as the calculations are made for an ideal compound.

TiMnSb, reported in the BeZrSi structure-type, is calculated more stable in the TiNiSi structure-type. This disagreement can easily be explained by the fact that TiMnSb has been experimentally obtained under extreme conditions [38] (high pressure and high temperature) and may be a metastable phase at ambient pressure. Besides, in this compound, Ti and Mn are assumed to be randomly distributed on both the *T* and *M* sites [38].

For CrCoAl reported as a half-Heusler compound by Luo *et al.* [39], a positive enthalpy of formation is obtained for the MgAgAs structure-type. In order to understand this disagreement, a sample of CrCoAl has been synthetized in this work, strictly following the experimental conditions of Luo *et al.* [39]. The obtained powder has been analysed by X-ray diffraction and all major peaks could be indexed in a cubic structure, corresponding to $Cr_{0.34}Co_{1.1}Al_{0.56}$ (*Pm-3m*, 221, CsCl structure-type). The superstructure peak characteristic of the ordering of the CsCl type into the MgAgAs type that should appear at 2θ = 27° is neither observed in our XRD pattern nor in the original article by Luo *et al.* [39]. So we can conclude that both our calculations and experiments prove that CrCoAl is not stable in the MgAgAs structure-type but rather in the CsCl prototype.

TiFeSn is calculated as stable in the TiNiSi type while it has been reported in the half-Heusler structure-type by Kuentzler *et al.* [40]. This sample was synthesized by arc-melting and annealed at 1123K during one week. However according to other reports, both theoretical and experimental [41, 42], in the isothermal section (773K and 873K) of the ternary phase diagram there is no observation of the equiatomic ternary compound TiFeSn. Moreover, DFT calculations [41] underline the instability of TiFeSn (in the cubic structure-type) which decomposes into other compounds of the phase diagram as $Ti_6Sn_5$, $Ti_2Sn_3$ and $TiFe_2Sn$. These studies of the phase diagram report the existence of TiFeSn only at high temperature (annealing at 1273K) but crystallizing in the LiGaGe structure-type. It is hence difficult to conclude about the most stable structure-type for TiFeSn, in particular at 0 K.

ZrCuSn has been reported by Skolozdra *et al.* [43] as crystallizing in the TiNiSi structure-type. However according to our calculations, this compound is more stable in the hexagonal BeZrSi structure-type than in the TiNiSi structure-type. When examining the crystallographic databases [44], the study of the parent compounds, TiCuSn and HfCuSn points to the fact that they do not crystallize in the TiNiSi structure-type but rather in the hexagonal LiGaGe one. In the case of TiCuSn, both the BeZrSi and LiGaGe types have been reported [45, 46]. Thus, in order to go beyond these results, we decided to perform additional calculations: the *T*-CuSn compositions have also been calculated in the LiGaGe structure-type and the results are presented in Table 2. First, it can be noticed that in each case, the values of the enthalpy of formation are close for at least two structure-types. In the case of TiCuSn and LaCuSn, it is impossible, given the uncertainty on the calculation (0.5kJ on $\Delta_f H$), to determine the most stable structure-type. However, in the case of ZrCuSn, the LiGaGe structure-type is calculated as the most stable type and the TiNiSi structure-type is calculated as the less stable.

*Table 2 : Value of the enthalpy of formation at 0K for T-CuSn compounds in 5 different structure-types.*

| Element *T*-Cu-Sn | $\Delta_f H$ (MgAgAs) (kJ/mol) | $\Delta_f H$ (TiNiSi) (kJ/mol) | $\Delta_f H$ (ZrNiAl) (kJ/mol) | $\Delta_f H$ (BeZrSi) (kJ/mol) | $\Delta_f H$ (LiGaGe) (kJ/mol) |
|---|---|---|---|---|---|
| Ti | -12.674 | **-21.040** | -13.465 | **-20.996** | **-21.032** |
| Zr | -28.619 | -16.296 | -29.715 | -36.661 | **-37.462** |
| Hf | -16.979 | -6.624 | -18.192 | -24.841 | **-25.815** |
| La | **-57.926** | **-57.664** | -47.736 | **-57.778** | **-57.784** |

Finally, NbMnSi has been reported in the ZrNiAl structure-type [47, 48], whereas the calculations predict the TiNiSi structure-type as more stable than the ZrNiAl structure-type. Even if an orthorhombic deformation has been noticed in the related NbMnGe and TaMnSi compounds [49], the crystallographic study of NbMnSi led to an indexation of all the diffraction peaks in a hexagonal structure. Moreover, no deviation from the stoichiometry has been reported in the literature [47-49]. In order to understand the disagreement between both calculations and experimental results, additional phonon calculations of NbMnSi in the TiNiSi and the ZrNiAl structure-type were performed in the present work (see SI). The results are consistent with the previous energy calculations: the phonon frequencies are real in the TiNiSi structure-type while imaginary frequencies branches are present in the ZrNiAl structure-type. Thus, according to our calculations, NbMnSi is mechanically unstable and should not crystallize in the ZrNiAl structure-type.

As a conclusion, the disagreement observed between literature and the calculations can be explained in every case except one and is not the expression of an error from the model, confirming the reliability of our DFT calculations. Indeed, if a 1:1:1 composition exists in one of the 4 calculated structure-types, our model can predict which one is the most stable with an excellent accuracy. Thus, the calculated most stable structure-type will be considered as the most probable one for the unknown compounds.

ii. Electronic structure

The first principles calculations give also information about the electronic structure of the screened compounds. In particular, the value of the density of states at the Fermi level is calculated in order to exclude the metallic compounds and used to identify the possible semiconductors among the 472 compounds. As already presented in Figure 3, the 21 compounds which are found as non-metallic and could therefore be interesting for thermoelectricity or other applications are reported in Table 3.

*Table 3 : Ternary intermetallic compounds which present a DOS at the Fermi level smaller than 0.5 states/eV with their calculated most stable structure-type and comparison with the literature data.*

| Compounds | Calc. Structure-type | Crystal databases |
|---|---|---|
| TiVAl | BeZrSi | Unreported |
| TiCoSb | MgAgAs | MgAgAs |
| TiNiSn | MgAgAs | MgAgAs |
| ZrCoSb | MgAgAs | MgAgAs |
| ZrNiSn | MgAgAs | MgAgAs |
| HfTiSn | BeZrSi | Unknown |
| HfCoP | TiNiSi | TiNiSi |
| HfCoSb | MgAgAs | MgAgAs |
| HfNiSn | MgAgAs | MgAgAs |
| HfCuSb | TiNiSi | Unreported |
| VFeSb | MgAgAs | MgAgAs |
| VCoSn | MgAgAs | Unreported |
| NbFeSb | MgAgAs | MgAgAs |
| NbCoSn | MgAgAs | MgAgAs |
| TaFeSb | MgAgAs | Unknown |
| TaCoSn | MgAgAs | MgAgAs |
| MoCoAl | MgAgAs | Unreported |
| WFeSi | MgAgAs | $MgZn_2$ |
| WCoAl | MgAgAs | $MgZn_2$ |
| LaCuSn | MgAgAs/BeZrSi/TiNiSi/LiGaGe | BeZrSi / LiGaGe |
| SrCuSb | BeZrSi | BeZrSi |

For the 21 compounds predicted as non-metallic, an additional research has been made particularly with regard to their stability. Indeed as explained in the methodology section, only the 1:1:1 compounds which are reported in the crystallographic database in one of the four calculated structure-types are classified as “agreed” in the Figure 3. However, additional information may be available for other compositions, in particular through the study of the ternary phase diagram. In the next part of this paper, each potential semiconductor is presented and the results are discussed.

*MgAgAs structure-type*

Over these 21 compounds predicted as semiconductors, 16 are calculated in the half-Heusler or MgAgAs structure-type. It can be noticed that all the half-Heusler compounds which are predicted as non-metallic follow the 18 electron rule [50]. 10 are already experimentally known in this structure-type and have already been reported in the literature as promising thermoelectric compounds, in good agreement with our prediction of a semiconducting ground state. The crystal structure of LaCuSn has already been discussed and only 5 presumable half-Heusler are not reported in the crystallographic databases: VCoSn, TaFeSb, MoCoAl, WCoAl and WFeSi. A more specific study of the literature shows that, WCoAl and WFeSi exist and are reported to crystallize in a pseudo-binary crystalline structure, e.g. the $MgZn_2$ structure-type in which *M* and *X* share the Zn positions [51, 52]. In the ternary Mo-Co-Al phase diagram, there is no reported equimolar compound [53]. The existence of VCoSn is controversial: it has been identified as crystallizing in the MgAgAs structure-type [54] while it does not exist according to Asaas *et al.* [55]. Only one semi-conducting composition is yet completely unknown: TaFeSb.

*TiNiSi structure-type*

Two of the 21 compounds predicted as semiconductors crystallize in the TiNiSi structure-type: HfCoP and HfCuSb. HfCuSb has never been reported, to our best knowledge, in any crystal database. However, a study of the ternary *Hf-Cu-Sb* phase diagram at 770K [56] has reported that the equiatomic ternary compound does not exist, similarly to the other ternary (Ti, Zr)-Cu-Sb systems [57]. On the contrary HfCoP has been reported as crystallizing in the TiNiSi structure-type [58, 59] but no measurement of its electronic or transport properties were presented. In the article of Kleinke *et al.* [59], the comparison of HfCoP with its parent compound HfNiP led the authors to expect a metallic conductivity as the calculated density of states at the Fermi level was not zero. As presented in Figure 4, our calculated density of states does not exhibit also a zero value to the Fermi level but this value is smaller than 0.5 states/eV by atom. Experiments are underway to determine the ground state of HfCoP.

*Hexagonal structure-types*

All the compounds calculated in the ZrNiAl prototype are predicted as metallic. Three compounds calculated in the BeZrSi structure-type are found as non-metallic: TiVAl, HfTiSn and SrCuSb. In the ternary Ti-V-Al phase diagram, only pseudo-binary compounds are reported [60]. To our best knowledge, there is no report on HfTiSn. In the case of SrCuSb, the crystallographic structure has been reported [61] and corresponds to the predicted one, however there is no measurements of its electronic or transport properties.

The last compound reported in the table is LaCuSn. However, there is some uncertainty on the crystal structure of this compound. Indeed, the lowest value of the enthalpy of formation is obtained for the MgAgAs structure-type in which the compound is found as non-metallic. However, as for ZrCuSn (Table 1), the enthalpy of formation is very similar for several structure-types (BeZrSi, TiNiSi and LiGaGe) and it is difficult to conclude about the most stable one, given the uncertainty of the DFT calculations. Over the four possible prototypes, the MgAgAs structure-type is the only non-metallic. Besides, looking at the reported data in the crystal databases, the hexagonal structure-types are the most likely as LaCuSn has been reported in two different structure-type: BeZrSi [62] and LiGaGe [63]. Thus, in view of these results it is probable that the cubic structure is not the most stable and therefore this compound is likely to be a metal.

## V. Conclusions

We investigated 2280 possible configurations in 4 different structure-types (TiNiSi, MgAgAs, BeZrSi and ZrNiAl) in order to identify the stable and semiconducting ones. Comparison with the available data in crystallographic databases allowed us to validate the robustness of our model since almost all the previously reported compounds have been calculated as stable in the correct structure-type. Over the 472 most stable compounds, only 21 have been found as non-metallic, most of them are half-Heusler (MgAgAs) and follow an 18 electrons rule. Among these 21 compositions, 10 are already well known. After a careful examination of their reported ternary phase diagram, 6 compounds among the 21 can be excluded as they are not stable compared to other ternary or pseudo-binary compounds or equilibrium between compounds. LaCuSn most likely crystallizes in a metallic hexagonal structure. Finally, 4 compositions with unknown electronic properties, with a non-reported existence (TaFeSb, HfTiSn) as well as with an already reported crystal structure (HfCoP, SrCuSb) should further be investigated as potential semiconductors and maybe thermoelectric materials.

## VI. Acknowledgements

DFT calculation were performed using HPC resources from GENCI-CINES (Grant 2017-096175).